\documentclass[a4paper,11pt]{article}
\usepackage{pos}
\title{
Calibration and Performance Validation of the SST-1M Telescopes Using Crab Nebula Observations }

\author*[a]{Thomas Tavernier}
\author[a]{Jakub Jury\v sek }
\author[a,b]{Vladimir Novotny }
\affiliation[a]{FZU - Institute of Physics of the Czech Academy of Sciences, \\
  Na Slovance 1999/2, Prague 8, Czech Republic}

\affiliation[b]{Faculty of Mathematics and Physics, Charles University,\\
V Hole\v sovi\v ckach 2, Prague 8, Czech Republic}

\onbehalf{on behalf of the SST-1M Collaboration\hyperref[authorlist]{*}} 

\emailAdd{tavernier@fzu.cz}

\abstract{SST-1M is a prototype single mirror Small Sized Cherenkov Telescope designed for very high energy (VHE) gamma-ray astronomy. With a 4 meter primary mirror and a 5.6 meter focal length, it provides a wide 9 degree optical field of view, optimized for detecting VHE gamma-rays from 1 TeV to several hundred TeV. Its focal plane is equipped with DigiCam, a fully digital trigger and readout camera made of 1296 silicon photomultiplier (SiPM) pixels. The use of SiPM sensors enables observation under high night sky background (NSB) conditions, significantly enhancing the instrument's duty cycle and allowing observations under moonlight.

Currently, two SST-1M telescopes are deployed at the Ond\v rejov Observatory in the Czech Republic, operating in stereo, at 510 m altitude, to observe astrophysical sources. This contribution presents the SiPM calibration procedure and performance validation of the instrument, based on updated results from Crab Nebula observations.
}

\ConferenceLogo{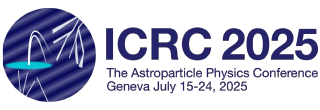}

\FullConference{39th International Cosmic Ray Conference (ICRC2025)\\
 15–24 July 2025\\
Geneva, Switzerland\\}

\begin{document}

\maketitle

\section{Introduction}
Single-Mirror Small-Size Telescopes (SST-1M) are two telescopes designed to observe atmospheric showers induced by gamma-ray with energies between 1 and 300 TeV.

The SST-1M telescope uses a Davies-Cotton design with a 5.6 m focal length. Its mirror consists of 18 spherical hexagonal facets (78 cm flat-to-flat) with an 11.2 m radius of curvature. The mirror area is 9.42 m², delivering an optical point spread function (PSF) of 0.09° on-axis and 0.21° at 4° off-axis. After accounting for shadowing and mirror reflectivity, the effective collecting area is 6.5 m². The aperture of the focal plane is composed of a 3 mm thick Borofloat window, which integrates a narrow-band optical filter composed of dielectric layers and an anti-reflective coating. A thorough description of the instrument and its components is given in \citep{Alispach_2025}.
 
The Cherenkov camera is equipped with 1296 silicon photomultipliers (SiPMs), offering resilience to high NSB. Two SST-1M prototypes are currently deployed at the Ondřejov Observatory in the Czech Republic, separated by 152.5 meters. Synchronized to nanosecond precision using the White Rabbit network, they are operating in stereoscopic mode and actively collecting gamma-rays data on astrophysical sources since their installation in 2022.

The Crab Nebula is a composite (SNR, PWN) TeV source, hosting a 33 ms pulsar. It was the first VHE gamma-ray source ever detected and is used as benchmark for VHE astronomy. Its spectrum extends to PeV energies \citep{LHAASO_2021} and, despite GeV variability, it shows a stable VHE flux. The first detection of Crab nebula by SST-1M telescopes was reported in \citep{the_sst-1m_collaboration_analysis_2024}, in this contribution, we assess the stereo performance of the SST-1M telescopes using Crab Nebula observations taken between September 2023 and March 2025.

\section{Timing validation using Crab pulsar in B-UVB band}

White Rabbit network \citep{white_rabbit} is used to synchronize both DigiCam clocks at nanosecond level and provide time reference for event timestamping with a precision of 16 ns. The system was deployed in April 2023 and connected to a GPS reference  March 10, 2024. To test the absolute timing accuracy of timestamps, we used optical to UV modulated signal from the Crab pulsar collected during standard observations.

The SiPMs are DC-coupled to monitor the current and correct for variations in NSB. This DC coupling also improves the ability to detect signals above the NSB level. The DC level is measured directly in the DigiCam FPGA, enabling real-time monitoring and dynamic adaptation of the trigger threshold to varying NSB conditions. Additionally, the DC level, estimated for 1024 samples of 4~ns, is stored for each event  to allow accurate baseline subtraction at the analysis level.
We used this baseline information, to estimate the modulation of the NSB with Crab pulsar rotation phase.

For each event, including both triggered and monitoring events, the baseline ADC level, was estimated over a 4.096~$\mu$s window for pixels within 0.15° of the Crab pulsar position. This method has the advantage of requiring no dedicated data acquisition setup but results in a very low effective duty cycle: from a total of 49.1 hours of observation after data quality selection, the effective exposure time sums up to only 337.3 s.

\begin{figure}
    \centering
    \includegraphics[width=0.5\linewidth]{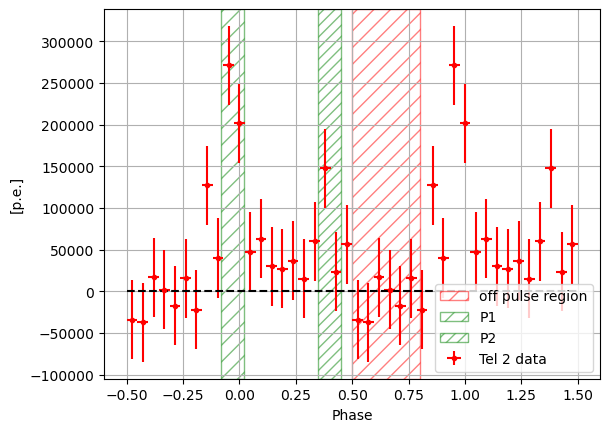}
    \caption{Modulation of p.e. above NSB as a function of the Crab pulsar rotational phase, measured by SST-1M-2. The data were obtained using baseline levels recorded in standard observation mode. The effective exposure time sums up to 337.3 s. The shaded region (phase 0.5–0.8) indicates the off-pulse window used as a reference. Two phases are represented.}
    \label{fig:psrcs2}
\end{figure}

The association of each baseline timestamp with the pulsar rotational phase was performed using monthly ephemerids provided by the Jodrell Bank Observatory \citep{mjbank}. Timing corrections, including barycentering, were applied using the PINT timing package \citep{Luo_2021,Susobhanan_2024}. The off-pulse level was estimated between the phase 0.5 to 0.8. The p.e. excess, above constant NSB, with the pulsar phase is shown in figure \ref{fig:psrcs2}. The obtained results show a clear pulse in coincidence with the position of the brightest optical pulse seen by other experiments.

\section{Analysis pipeline}

The \texttt{sst1mpipe} \citep{alispach_observation_2025,sst1mpipe_073} analysis pipeline is designed to produce high-level (DL3) data products in the Gamma-ray Astronomy Data Format (GADF), starting from the raw waveforms generated by the instrument. The pipeline is built upon the \texttt{ctapipe} \citep{karl_kosack_2021_5720333} framework and includes all analysis steps, such as SiPM response calibration, charge reconstruction, event image cleaning, gamma/hadron separation, and the reconstruction of the primary particle's energy and direction using random forests trained on Monte Carlo (MC) simulations. Detailed description of each analysis step are described in \citep{alispach_observation_2025}.

MC test dataset were also used to generates full enclosure Instrument Response Functions (IRFs) following GADF standards and using \texttt{pyirf} \citep{pyirf-icrc-2023}. These data products can be used to produce higher-level analyses, such as sky maps and energy spectra, using the \texttt{gammapy} \cite{axel_donath_2021_5721467} framework.

\section{Available dataset and quality selection}
We present in this contribution the analysis of Crab observations campaigns conducted during the winters 2023–2024 (43.4 h ) and 2024–2025 (84.0 h) for a total of 127.4 h of data before quality selection. Data have been collected using wobble pointing strategy for four different pointing offsets: 0.7° (19.0 h), 1.4° (95.7 h), 2.1° (18.3 h) and 3.1° (15.7 h). Observations were taken with both telescopes triggering independently. For stereo reconstruction, events from the same shower are matched using timestamps.

\begin{figure}
    \centering
    \includegraphics[width=0.49\linewidth]{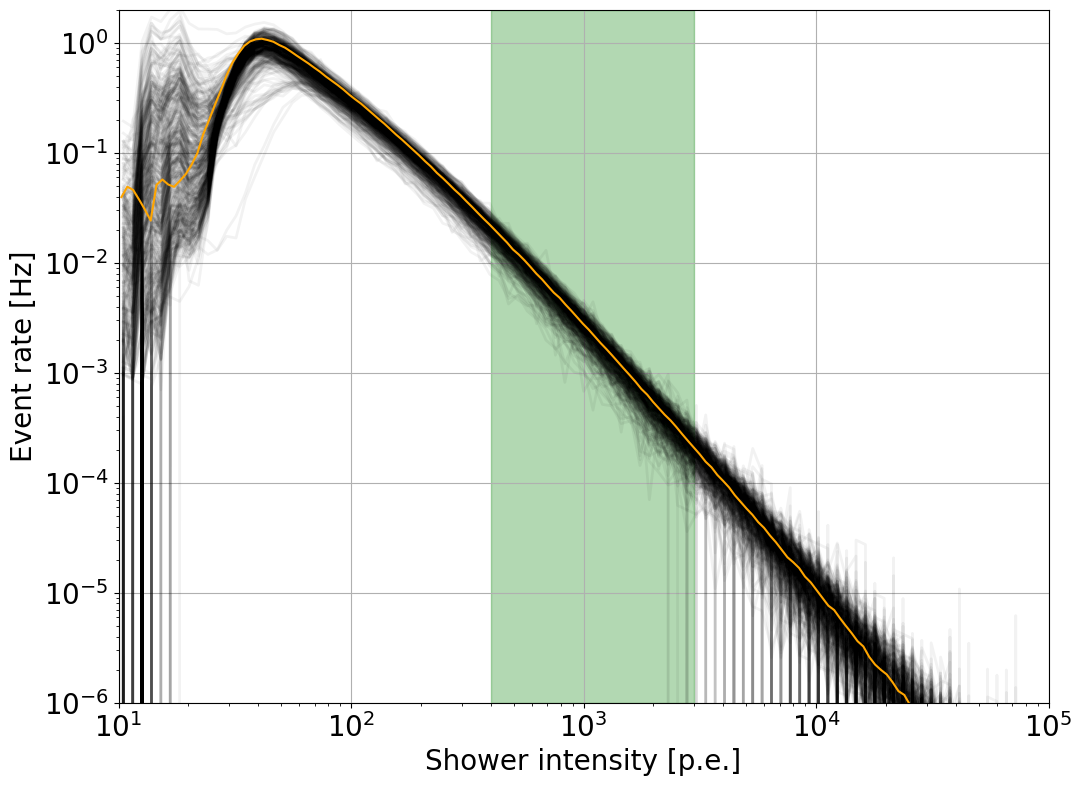}
    \includegraphics[width=0.45\linewidth]{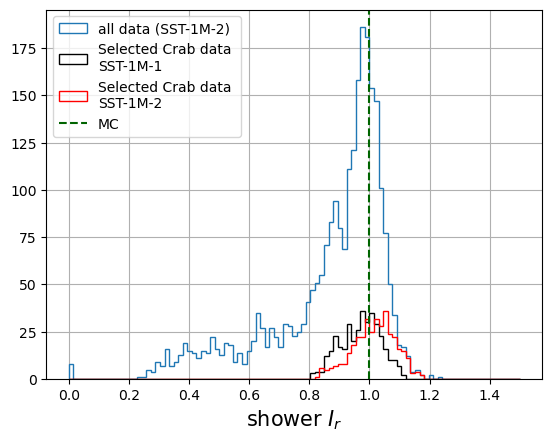}
    \caption{\textit{left} : Hadronic background event rate as a function of shower intensity for selected observation runs (black lines). Orange line indicates the MC rate used for reference. Both MC and real observation are corrected from zenithal angle dependence. The  green region highlights the intensity range (400–3000 p.e.) used for the extinction estimation.\\
    \textit{right} : Distribution of the light intensity ratio $I_r$ derived from hadronic background rates. The blue histogram shows all data taken with SST-1M-2. The black and red histograms correspond to selected Crab observations from SST-1M-1 and SST-1M-2, respectively. Only data with $I_r$ within 20\% of the MC reference (vertical green dashed line) are kept for analysis.}
    \label{fig:DQ}
\end{figure}

Data quality is assessed using the hadronic background rates before the application of the gammaness selection. The behavior of the hadron rate as a function of shower intensity ($I_\mathrm{sh}$), in the range of 400 to 3000 p.e., is modeled using a simple power-law function derived from MC simulations : 
$$R(I_\mathrm{sh})=A*I_\mathrm{sh}^{-B}$$ where $R$ 
is the event rate.  To estimate the atmospheric extinction factor in the data, we fit the light intensity ratio, $I_r$, in a scaled model : $R(I_\mathrm{sh}I_r)$ which have been corrected for zenith angle dependence. The resulting fit provides an estimation of atmospheric extinction relative to the MC. This procedure is used to reject observations affected by significant extinction as well as observations affected by clouds or technical issues. In practice, only observations with less than a 20\% difference  in hadron rates compared to the MC are retained for further analysis, effectively reducing the systematic errors induced by the atmosphere. This procedure ensures that the selected data and Monte Carlo simulations correspond to similar observational conditions, and it allows for the estimation of atmospheric extinction. The distribution of this intensity ratio for all observations and the selected subset is shown in figure \ref{fig:DQ}. Additionally, observations with important fluctuation of the baselines induced by high NSB are also excluded.
After data quality selection, the full dataset amounts to a total of $92.6$ hours of good quality observation.

\section{Results}
The selected dataset was used to evaluate the performance of the analysis pipeline by comparing the distributions of key parameters reconstructed using random forest algorithms in the excess measured on the Crab region and the one in MC point gamma simulations.

\subsection{Gamma hadron separation}
\begin{figure}[h]
    \centering
    \includegraphics[width=0.45\linewidth]{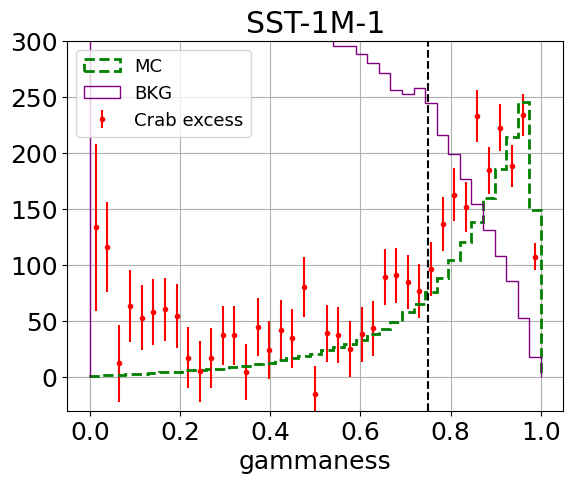}
    \includegraphics[width=0.45\linewidth]{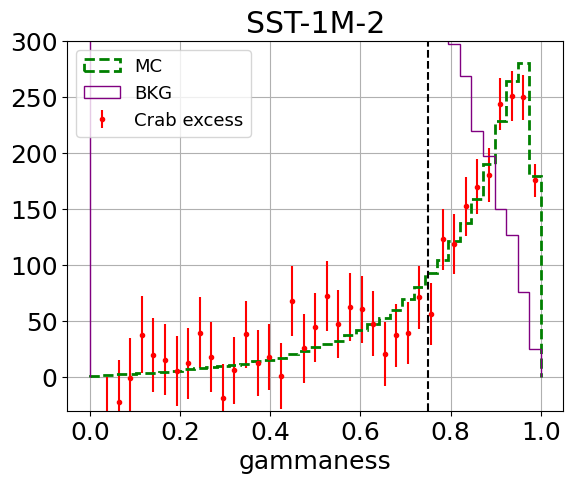}
    \caption{Distribution of gammaness for SST-1M-1 (\textit{left}) and SST-1M-2 (\textit{right}). The red markers represent the excess events from 0.2° region centered on the Crab Nebula after background subtraction. The green dashed line shows the distribution from MC gamma-ray simulations weighted on the Crab Nebula spectrum. The purple histogram corresponds to the gammaness distribution for hadronic background, measured on the off-regions. The vertical dashed line indicates the analysis cut applied for gamma/hadron separation.}
    \label{fig:gammaness}
\end{figure}

Gamma/hadron separation was performed using random forest classifier trained on MC simulation resulting in a gammaness parameter, between 0 and 1, that quantifies how gamma-like an event is. 
Figure \ref{fig:gammaness} shows the distribution of gammaness of the excess on the Crab Nebula for data collected at a zenith angle close to 30°. The background was subtracted using reflected region method, with an ON region of 0.2° radius centered on the Crab Nebula and 5 OFF regions, independently of the wobble offset of the observation.

\subsection{Crab Nebula spectrum}
\begin{figure}
    \centering
    \includegraphics[width=0.6\linewidth]{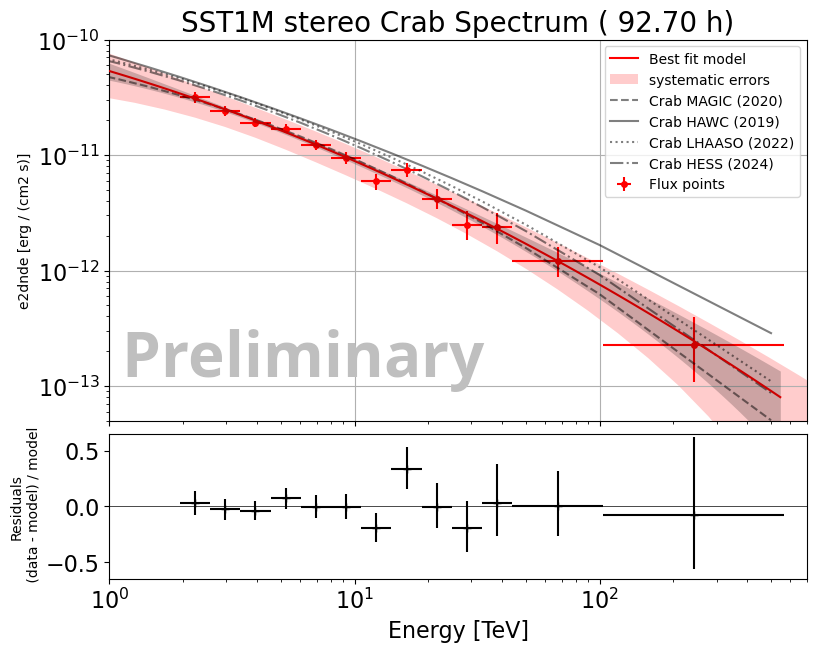}
    \caption{SED of the Crab Nebula measured with SST-1M, derived using a forward-folding likelihood fit in the energy range [2–1000] TeV. The red markers estimations of the flux points with statistical uncertainties, while the solid red line shows the fit using log parabola model. The shaded red band corresponds to the estimated systematic uncertainties. For comparison, measurements from MAGIC(2015) \citep{Aleksi__2015}, HAWC (2019)\citep{Abeysekara_2019}, LHAASO (2022)\citep{LHAASO_2021}, and H.E.S.S. (2024)\citep{Aharonian_2024} are overlaid.}
    \label{fig:SED}
\end{figure}

We conducted a spectral analysis on the dataset selected, using the \texttt{gammapy} framework. We used the reflected region method for background estimation, where the background is estimated from regions with the same offsets from the center of the FoV as the ON region.
For the source position, we adopted the coordinates of the Crab Nebula reported by \citep{Aharonian_2024}: $\alpha_{2000} = 83.629^\circ$, $\delta_{2000} = +22.012^\circ$. The on source region was defined with a $\theta$ cut of $0.12^\circ$, corresponding to approximately 80\% containment for point-like gamma-ray sources. 
We measure an excess of 976.5 events on the ON region (1172 counts, 195.4 estimated background), leading to more than 40 Li\&Ma $\sigma$ significance.  
The spectral energy distribution (SED) shown in figure \ref{fig:SED} was derived using forward-folding likelihood fit in the energy range 1-1000 TeV in \texttt{gammapy} framework. We assumed a log parabola spectral shape for the differential flux:
$$\Phi(E) = \Phi_0 \left( \frac{E}{E_0} \right)^{-\alpha - \beta \mathrm{log}\left( \frac{E}{E_0} \right)} $$

The reference energy is fixed at $E_0=$10 TeV. At this energy, the fitted differential flux is $\Phi_0 = (5.5 \pm 0.3_\mathrm{stat}) \times 10^{-14}$ cm$^{-2}$s $^{-1}$ TeV$^{-1}$. Fitted values for parameters of the log parabola are $\alpha = 2.92 \pm 0.05_\mathrm{stat}$ and $\beta = 0.06 \pm 0.04_\mathrm{stat}$.

A likelihood ratio test with a simple power-law model shows no significant evidence (2$\sigma$) for curvature in the spectrum.  Nevertheless, we chose to model the spectrum with a log parabola shape to facilitate comparison with measurements from other experiments. The obtained spectrum show good agreement with measurements from other gamma-ray experiments. Notably, 12 events in spatial coincidence with the Crab Nebula  were reconstructed above 50 TeV,  compared to an expected background of 2.3 events. This results in a 4.4 $\sigma$ Li\&Ma significance.

\begin{figure}
    \centering
    \includegraphics[width=0.45\linewidth]{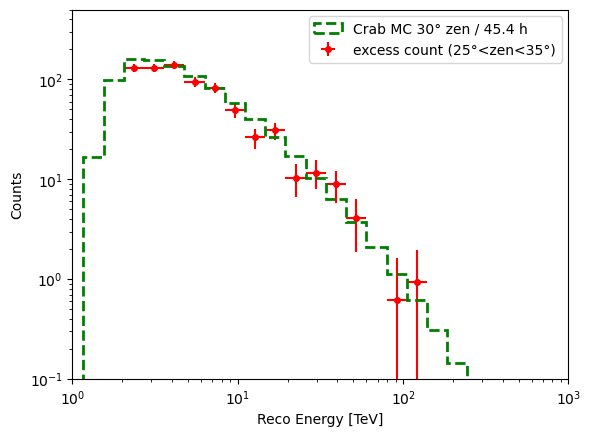}
    \includegraphics[width=0.45\linewidth]{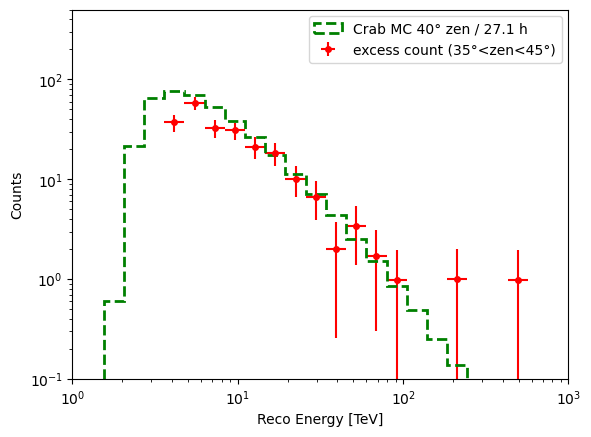}
    \caption{Distribution of reconstructed energies for both the excess measured at the Crab Nebula position (red points) and the Monte Carlo simulation of the Crab spectrum (dashed green histogram). Two zenith angles are shown : 30° (left) and 40° (right).}
    \label{fig:Energy}
\end{figure}

Additionally, Figure \ref{fig:Energy} presents the distribution of stereo reconstructed energies for both the excess measured at the Crab Nebula position and the MC simulation of the Crab spectrum for two zenith angles (30° and 40°). 

\subsection{Morphology and PSF}
\begin{figure}
    \centering
    \includegraphics[width=0.45\linewidth]{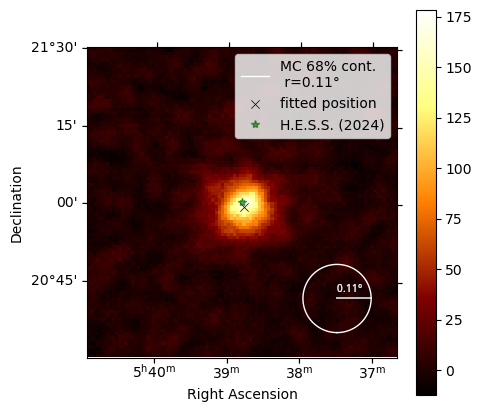}
    \includegraphics[width=0.45\linewidth]{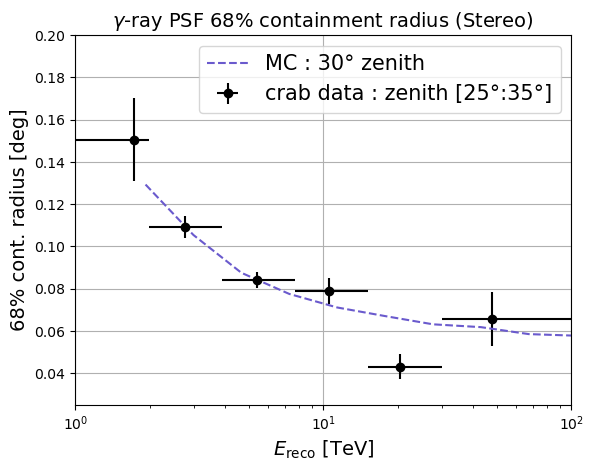}
    \caption{\textit{left:} Excess map of the Crab Nebula region obtained with the SST-1M telescopes, oversampled with a 0.03° disk kernel. The black cross marks the best-fit position of the gamma-ray excess from our data, while the green star indicates the Crab Nebula position as measured by H.E.S.S \citep{Aharonian_2024}.\\ 
    \textit{right :} Energy dependence of the 68\% containment radius of a fitted gaussian PSF  in stereo reconstruction. Black points represent values obtained from Crab Nebula data in the zenith angle range [25°–35°], while the dashed blue line corresponds to MC simulations of point-like gammas weighted on the Crab Nebula spectrum at 30° zenith. Error bars include statistical uncertainties only.
    }
    \label{fig:sky_dist}
\end{figure}

The excess and significance distributions of the Crab Nebula observations were computed using the ring background method \citep{puhlhofer_technical_2003} implemented in the \texttt{gammapy} framework. A background ring with a radius of 1$^\circ$ and a width of 0.3$^\circ$ was used to estimate background events, a $0.3^\circ$ region around the Crab Nebula was excluded from the background estimation. The excess distribution, convolved with a disk kernel of $0.03^\circ$ is shown in Figure \ref{fig:sky_dist}.

 We fitted a two dimensional, symmetrical Gaussian on the excess distribution to estimate the position of the excess. The best-fit coordinates were found to be $(\alpha_{2000} = 83.62^\circ, \delta_{2000} = 21.99^\circ ) \pm 0.01_\mathrm{stat}^\circ$. This position is consistent with the expected Crab Nebula location and in line with the pointing precision of the instrument measured to be $0.02_\mathrm{sys}^\circ$ \citep{Alispach_2025}.

Crab Nebula extension have been estimated by H.E.S.S. \citep{Aharonian_2024} to be approximately 1 arcminute in the [3-100] TeV range. We performed a 3D likelihood fit to test for a possible extension in our dataset. No significant extension was found in our analysis.
Assuming the Crab Nebula is compatible with a point-like source in our data, we fitted the $\theta^2$ distribution of the excess using a Gaussian PSF model for different energy bands. The same procedure was applied to MC simulations. The resulting 68\% containment radius of the fitted gamma-ray PSF is shown in Figure \ref{fig:sky_dist}.

\subsection{Conclusion}

We have presented the analysis of Crab Nebula using data collected over two observation campaigns with SST-1M Telescopes. Theses data were used to extensively test the performances of both the instrument and the data analysis pipeline from the calibration to high level data products.  Key parameters reconstructed by the random forests, including gammaness for the gamma/hadron discrimination, energy reconstruction, and gamma-ray PSF, shows good agreement with MC simulations. The spectral measurements of the Crab Nebula obtained by SST-1M are in good agreement with previous observations by established VHE gamma-ray instruments, validating the instrument's energy scale and sensitivity across the multi-TeV range.

The successful detection of the Crab pulsar optical pulses allowed us to confirm the precise timing provided by the White Rabbit network, paving the way for gamma-ray pulsar analysis. 

Finally, the results presented in this study demonstrate that the SST-1M telescope and its analysis chain are operational and suitable for scientific observations in the VHE gamma-ray domain.

{\scriptsize
\section*{acknowledgment}
This publication was created as part of the projects funded in Poland by the Minister of Science based on agreements number 2024/WK/03 and DIR/\-WK/2017/12. The construction, calibration, software control and support for operation of the SST-1M cameras is supported by SNF (grants CRSII2\_141877, 20FL21\_154221, CRSII2\_160830, \_166913, 200021-231799), by the Boninchi Foundation and by the Université de Genève, Faculté de Sciences, Département de Physique Nucléaire et Corpusculaire. The Czech partner institutions acknowledge support of the infrastructure and research projects by Ministry of Education, Youth and Sports of the Czech Republic (MEYS) and the European Union funds (EU), MEYS LM2023047, EU/MEYS CZ.02.01.01/00/22\_008/0004632, CZ.02.01.01/00/22\_010/0008598, Co-funded by the European Union (Physics for Future – Grant Agreement No. 101081515), and Czech Science Foundation, GACR 23-05827S.
}
\bibliographystyle{JHEP}
{\footnotesize

\bibliography{references}}

\clearpage
\section*{Full Authors List: SST-1M Collaboration}\label{authorlist}
\scriptsize
\noindent
C.~Alispach$^1$,
A.~Araudo$^2$,
M.~Balbo$^1$,
V.~Beshley$^3$,
J.~Bla\v{z}ek$^2$,
J.~Borkowski$^4$,
S.~Boula$^5$,
T.~Bulik$^6$,
F.~Cadoux$^`$,
S.~Casanova$^5$,
A.~Christov$^2$,
J.~Chudoba$^2$,
L.~Chytka$^7$,
P.~\v{C}echvala$^2$,
P.~D\v{e}dic$^2$,
D.~della Volpe$^1$,
Y.~Favre$^1$,
M.~Garczarczyk$^8$,
L.~Gibaud$^9$,
T.~Gieras$^5$,
E.~G{\l}owacki$^9$,
P.~Hamal$^7$,
M.~Heller$^1$,
M.~Hrabovsk\'y$^7$,
P.~Jane\v{c}ek$^2$,
M.~Jel\'inek$^{10}$,
V.~J\'ilek$^7$,
J.~Jury\v{s}ek$^2$,
V.~Karas$^{11}$,
B.~Lacave$^1$,
E.~Lyard$^{12}$,
E.~Mach$^5$,
D.~Mand\'at$^2$,
W.~Marek$^5$,
S.~Michal$^7$,
J.~Micha{\l}owski$^5$,
M.~Miro\'n$^9$,
R.~Moderski$^4$,
T.~Montaruli$^1$,
A.~Muraczewski$^4$,
S.~R.~Muthyala$^2$,
A.~L.~Müller$^2$,
A.~Nagai$^1$,
K.~Nalewajski$^5$,
D.~Neise$^{13}$,
J.~Niemiec$^5$,
M.~Niko{\l}ajuk$^9$,
V.~Novotn\'y$^{2,14}$,
M.~Ostrowski$^{15}$,
M.~Palatka$^2$,
M.~Pech$^2$,
M.~Prouza$^2$,
P.~Schovanek$^2$,
V.~Sliusar$^{12}$,
{\L}.~Stawarz$^{15}$,
R.~Sternberger$^8$,
M.~Stodulska$^1$,
J.~\'{S}wierblewski$^5$,
P.~\'{S}wierk$^5$,
J.~\v{S}trobl$^{10}$,
T.~Tavernier$^2$,
P.~Tr\'avn\'i\v{c}ek$^2$,
I.~Troyano Pujadas$^1$,
J.~V\'icha$^2$,
R.~Walter$^{12}$,
K.~Zi{\c e}tara$^{15}$ \\

\noindent
$^1$D\'epartement de Physique Nucl\'eaire, Facult\'e de Sciences, Universit\'e de Gen\`eve, 24 Quai Ernest Ansermet, CH-1205 Gen\`eve, Switzerland.
$^2$FZU - Institute of Physics of the Czech Academy of Sciences, Na Slovance 1999/2, Prague 8, Czech Republic.
$^3$Pidstryhach Institute for Applied Problems of Mechanics and Mathematics, National Academy of Sciences of Ukraine, 3-b Naukova St., 79060, Lviv, Ukraine.
$^4$Nicolaus Copernicus Astronomical Center, Polish Academy of Sciences, ul. Bartycka 18, 00-716 Warsaw, Poland.
$^5$Institute of Nuclear Physics, Polish Academy of Sciences, PL-31342 Krakow, Poland.
$^6$Astronomical Observatory, University of Warsaw, Al. Ujazdowskie 4, 00-478 Warsaw, Poland.
$^7$Palack\'y University Olomouc, Faculty of Science, 17. listopadu 50, Olomouc, Czech Republic.
$^8$Deutsches Elektronen-Synchrotron (DESY) Platanenallee 6, D-15738 Zeuthen, Germany.
$^9$Faculty of Physics, University of Bia{\l}ystok, ul. K. Cio{\l}kowskiego 1L, 15-245 Bia{\l}ystok, Poland.
$^{10}$Astronomical Institute of the Czech Academy of Sciences, Fri\v{c}ova~298, CZ-25165 Ond\v{r}ejov, Czech Republic.
$^{11}$Astronomical Institute of the Czech Academy of Sciences, Bo\v{c}n\'i~II 1401, CZ-14100 Prague, Czech Republic.
$^{12}$D\'epartement d'Astronomie, Facult\'e de Science, Universit\'e de Gen\`eve, Chemin d'Ecogia 16, CH-1290 Versoix, Switzerland.
$^{13}$ETH Zurich, Institute for Particle Physics and Astrophysics, Otto-Stern-Weg 5, 8093 Zurich, Switzerland.
$^{14}$Institute of Particle and Nuclear Physics, Faculty of Mathematics and Physics, Charles University, V Hole\v sovi\v ck\' ach 2, Prague 8, Czech~Republic.
$^{15}$Astronomical Observatory, Jagiellonian University, ul. Orla 171, 30-244 Krakow, Poland.

\end{document}